\documentclass[10pt]{article}
\usepackage{amsfonts}
\usepackage{color,caption}
\usepackage{graphicx}
\usepackage{amsmath}
\usepackage{float}
\usepackage{amssymb}
\providecommand{\U}[1]{\protect\rule{.1in}{.1in}}
\textwidth=6.5in \hoffset=-.75in \voffset=-1in \numberwithin{equation}{section}
\numberwithin{figure}{section}

\newcommand {\be}{\begin{equation}}
 \newcommand {\ee}{\end{equation}}
 \newcommand {\bea}{\begin{eqnarray}}
 
 \newcommand {\eea}{\end{eqnarray}}
 
\def \th {\theta}
\def \O {\Omega}
\def \f {\frac}

\def \D {\Delta}

\begin{document}
\begin{titlepage}
\bigskip \begin{flushright}
\end{flushright}
\vspace{1cm}
\begin{center}
{\Large \bf {Deformed conformal symmetry of Kerr-Newman-NUT-AdS black holes}}\\
\vskip 1cm
\end{center}
\vspace{1cm}
\begin{center}
M. F. A. R. Sakti$^{a,b,}${\footnote{mus561@usask.ca}}, A. M.
Ghezelbash$^{a,}${\footnote{amg142@campus.usask.ca}}, A. Suroso$^{b,c,}${\footnote{agussuroso@fi.itb.ac.id}}, F. P. Zen$^{b,c,}${\footnote{fpzen@fi.itb.ac.id}}
\\
$^a $Department of Physics and Engineering Physics, University of Saskatchewan,
Saskatoon, Saskatchewan S7N 5E2, Canada\\
$^b $Theoretical Physics Lab., THEPI Division, Institut Teknologi Bandung, Jl. Ganesha 10, Bandung, 40132, Indonesia \\
$^c $Indonesia Center for Theoretical and Mathematical Physics (ICTMP), Institut Teknologi Bandung, Jl. Ganesha 10 Bandung, 40132, Indonesia \\
\vspace{1cm}
\end{center}

\begin{abstract}
We find generators of the conformal symmetry for the class of Kerr-Newman-NUT-AdS black holes from the deformed scalar probe equation. We find two classes of solutions for the generators (also known as conformal $ J $ and $ Q $ pictures). The two classes of deformed generators are the extension of similar generators for the regular conformal symmetry. Moreover, we find that the two pictures can be generalized and extended into a general picture.  In each picture, the generators produce an extended local family of $SL(2,\mathbb{R})_L \times SL(2,\mathbb{R})_R$ hidden conformal symmetries for the Kerr-Newman-NUT-AdS black holes which are 
parameterized by one deformation parameter.  
We find the absorption cross-section of the scalar probes for the Kerr-Newman-NUT-AdS black holes, which in turn, supports the existence of Kerr/CFT correspondence. Moreover, our deformed conformal generators for the Kerr-Newman-NUT-AdS black holes provide the deformed conformal generators for the non-rotating Reissner-Nordstr{\"o}m-NUT-AdS black holes.

\end{abstract}
\end{titlepage}\onecolumn
\bigskip

\section{Introduction}

The well-known Kerr/CFT correspondence states a relation between the associated physical quantities of the extremal 
(and also extended to non-extremal) four-dimensional Kerr black hole and almost similar physical quantities 
in a conformal field theory (CFT) \cite{stro, Compere}.
The generalization of the correspondence to the other extremal (as well as non-extremal) rotating black holes, 
in both four dimensions, and higher than four dimensions, have been extensively considered, during the last decade \cite{KerrCFT1}-{\cite{Sakti2019b}}. 

To find the correspondence for the non-extremal rotating black hole, we may look at the solution space of a probe scalar field in the background of the non-extremal rotating black hole which is known as the hidden conformal symmetry \cite{Castro}. 

\textcolor{black}{We note that any black hole can have two asymptotes at the large distance  and the near horizon. The large distance behaviour and the associated symmetry group is different for flat, dS and AdS cases. }
We should emphasize that for the extremal rotating black holes, the near horizon geometry of the black holes is always an \textcolor{black}{AdS-like geometry},
and so the existence of the conformal symmetry is guaranteed in the light of AdS/CFT correspondence. 
For the  non-extremal black holes, the near horizon geometry is not an AdS-like geometry. In fact, it is a Rindler geometry, \textcolor{black}{which is a flat metric in a non-inertial coordinates. 
}
\textcolor{black}{It is worth noting that any scalar field theory on the background of a flat spacetime, can be expressed as a conformal field theory. So it would not be surprising that the conformal symmetry can emerge in the background of near horizon geometry of a non-extremal rotating black hole.}

\textcolor{black}{We also stress that in this article, by any reference to the holographic duality between the rotating black hole and the CFT, we exactly mean that the wave equation of a scalar probe field, on the near horizon geometry of the rotating black hole, possesses the conformal symmetry \cite{KerrCFT1}-{\cite{Castro}}.}

We try explicitly to find the conformal symmetry by looking at the scalar wave equation (or other wave equations with higher spins), \textcolor{black}{
 in the near horizon geometry of the non-extremal black holes}. This is the main reason, to call the symmetry as the hidden conformal symmetry.  The conformal generators are obtained by matching the Casimir operators of  the conformal algebra $SL(2,\mathbb{R})_L \times SL(2,\mathbb{R})_R$, to the scalar (or higher spin fields) wave equation in the background of the non-extremal black holes. In references \cite{othe1}-\cite{KSChen}, the authors successfully construct the $SL(2,\mathbb{R}) \times SL(2,\mathbb{R})$ algebra, in such a way that the quadratic Casimir operators of the algebra generate exactly the wave equation of the probe field. It is worth noting that $SL(2,\mathbb{R}) \times SL(2,\mathbb{R})$ symmetry is also the symmetry of AdS$ _3 $ space. 

Moreover for some charged rotating black holes, there is more than one dual hidden CFT. As an example,\textcolor{black}{ for the four-dimensional Kerr-Newman black hole, there are two dual hidden CFTs, associated with the solution space of the scalar wave equation}. The first CFT is associated to the rotation of black hole, while the second CFT is associated to the charge of black hole. The former and latter CFTs are called CFTs in $J$ and $Q$ pictures, respectively \cite{Chen-2fold}-\cite{Chen-genhidden4d5d}. \textcolor{black}{A counterexample in which there is only one CFT associated with the scalar wave equation, is the rotating charged  Kerr-Sen black hole}. The four-dimensional Kerr-Sen solution is an exact solution in the low energy limit of the
heterotic string theory \cite{sen}. The solution includes a black hole in the presence of three fields; a dilaton field, an antisymmetric second-rank-tensor field and the Maxwell's field \cite{sen}. The Kerr-Sen black hole is dual, only to one CFT in $J$ picture  \cite{Ghezelbash}. One other interesting feature of the Kerr-Sen black hole is that the generators of the hidden conformal symmetry in $J$ picture, can be extended to a new set of generators, by including a deformation parameter in the wave equation for the probe field \cite{Lowe}.  One important advantage of the deformed hidden conformal symmetry is that it provides the hidden conformal symmetry for the non-rotating black holes such as Schwarzschild black hole \cite{Lowe} or Gibbons-Maeda-Garfinkle-Horowitz-Strominger black hole \cite{Ghezelbash}.

\textcolor{black}{Inspired by the existence of two dual hidden conformal symmetries for the scalar field in the background of the Kerr-Newman black holes }and also the interesting invariance of the Kerr-Newman-NUT spacetime under the transformation of mass to the NUT charge and radius to the angular coordinate \cite{INdi1, INdi2}, we investigate the existence of the dual hidden conformal symmetry(ies) for the scalar wave equation in the background of the  Kerr-Newman-NUT-AdS geometries. 

The paper is organized as follows.  In section \ref{gen-hidden}, we consider the wave equation for an scalar probe in the background of the Kerr-Newman-NUT-AdS spacetime and show that the wave equation separates completely.
In section \ref{deformed-hiddensym}, we first find the generators of the deformed conformal symmetry in the $J$ picture. We then consider the deformed wave equation for the scalar probe and find the generators of the deformed conformal symmetry in the $Q$ picture. We also find the deformed right and left temperatures and the central charges of the dual CFTs. These quantities lead to the microscopic entropy for the dual CFTs,  in perfect agreement with the macroscopic Bekenstein-Hawking entropy. After that, we consider the extension of former two pictures and find the generators of the deformed conformal symmetry in the general picture. We also discuss the deformed hidden conformal symmetry in the $J$ and $Q$ pictures as the limits of results in the general picture. In section \ref{scat}, we consider the scattering of the charged scalar fields in the background of the Kerr-Newman-NUT-AdS black holes. We find the scattering cross-section of the charged scalar field, based on the deformed wave equation for the scalar probe in the general picture. We also discuss the proper limits of the scattering results in the general picture to find the results in the $J$ and $Q$ pictures. We get enough support for the correspondence between the Kerr-Newman-NUT-AdS black holes and the dual CFTs in the different pictures.  In section \ref{deformedRNNUTAdS}, we consider the non-rotating Reissner-Nordstr{\"o}m-NUT-AdS black holes as the limit of the Kerr-Newman-NUT-AdS black holes and present the generators of the deformed hidden conformal symmetry. We conclude the paper with conclusions in section \ref{conc}.

\section{The wave equation for the charged scalar field in the background of the Kerr-Newman-NUT-AdS black holes}\label{gen-hidden}

In this section, we consider the wave equation for a charged massless scalar probe on the geometry of a Kerr-Newman-NUT-AdS black hole, which is solution to the Einstein-Maxwell theory with non-vanishing cosmological constant \cite{Chen-2fold,Demianski1976,Griffiths2005,Podolsky2009,Chen-genhidden4d5d}. The Kerr-Newman-NUT-AdS geometry is represented by the following metric
\begin{eqnarray}
ds^2 &=& - \frac{\Delta_r}{\Xi^2\varrho^2} \left[ dt \!-\! \{a \sin^2\theta + 2n(1-\cos\theta) \} d\phi \right]^2 +  \frac{\varrho ^2}{\Delta_r}dr^2 + \frac{\varrho ^2}{\Delta_\th} d\theta^2 \nonumber\\ 
& & + \frac{\Delta_\theta \sin^2\theta}{\Xi^2 \varrho ^2}\left[adt - \{r^2+(a+n)^2 \} \right] d\phi^2,\label{KNNUTAdSmetric} \
\end{eqnarray}
where
\be
\varrho^2=r^2+(n+a\cos\th)^2 ,\label{rho}\ee
\be
\Delta_r=r^2- 2M r+e^2 +g^2 {+} \frac{r^2(r^2+6n^2+a^2)}{l^2}\textcolor{black}{+}\frac{(3n^2-l^2)(a^2-n^2)}{l^2} ,\label{Del}\ee
\be
\D_\th =1\textcolor{black}{-}\frac{a\cos\th}{l^2}(4n+a\cos\th), \label{Del2}\ee
\be
\Xi= 1\textcolor{black}{-} \frac{a^2}{l^2}, ~ \Lambda = \textcolor{black}{-}\frac{3}{l^2}.
\label{deltainkerr}
\ee
In (\ref{KNNUTAdSmetric}), $M$ and $a$ are mass and rotational parameters of the black hole, $e$ and $g$ are electric and magnetic charges of the black hole and $n$ is the NUT parameter. Moreover, $\Lambda $ is the cosmological constant. The corresponding electromagnetic potential is given by \cite{Podolsky2009}
\begin{eqnarray}
A_\mu dx^\mu &=& \frac{-er\left[a dt- \left( (a+n)^2 - (n+a \text{cos}\theta)^2 \right) d\phi \right]}{a\varrho ^2 \Xi}  \nonumber\\
&-& \frac{g(n+a\text{cos}\theta)\left[a dt- \left( r^2 + (a+n)^2 \right) d\phi \right]}{a\varrho ^2 \Xi} .\ \label{eq:electromagneticpotential}
\end{eqnarray}
The non-vanishing NUT charge $n$ in (\ref{KNNUTAdSmetric}) can be interpreted as the gravitational analog of a magnetic charge \cite{ME,ME2,SaktiZen}. The coordinate $t$ parametrizes a circle $S^1$ fibered over the non-vanishing sphere parametrized by $(\theta,\phi)$, where the rotation parameter and the black hole charges are zero.  The coordinate $t$ has a periodicity of $\pi n$ to avoid any conical singularities \cite{ME,ME2}. In fact, the geometry of a constant-$t$ surface is that of a Hopf fibration of $S^1$ over an $S^2$, and the metric (\ref{KNNUTAdSmetric}) describes the contraction or expansion of this 3-sphere in spacetime regions. There are no closed timelike curves and the spacetime is asymptotically locally Anti-de Sitter.

The Bekenstein-Hawking entropy, Hawking temperature, angular velocity and the electric potential on the horizon of the black hole (\ref{KNNUTAdSmetric}) are given by
\begin{eqnarray} \label{en}
&&S_{BH}=\textcolor{black}{\frac{\text{Area}}{4}}=\f{\pi}{\Xi} \left[r_+^2+(a+n)^2\right],\label{en1}\\
&&T_H=\f{2(r_+-M)l^2 \textcolor{black}{+}2r_+(2r_+^2+6n^2+a^2)}{4\pi(r_+^2+(a+n)^2)l^2 \Xi},\label{en2}\\
&&\O_H=\f{a}{r_+^2+(a+n)^2},\label{en3}\\
&&\Phi_H=\f{er_+}{\left[r_+^2+(a+n)^2\right]\Xi},\label{en4}
\end{eqnarray}
respectively, where $r_+$ is the outer horizon of the black hole. \par
We consider a massless charged scalar probe \footnote{\textcolor{black}{The probe field is a massless field which does not induce any 
back-reaction on the background black holes. 
For a massive scalar field, there is an extra term $ \mu^2 \rho^2 $ in the scalar wave equation, where $ \mu $ is the mass of scalar field \cite{
Vieira2014}.  The presence of the mass term hinders to establish the holography between the black holes and the CFT.}}, in the background of the Kerr-Newman-NUT-AdS black hole (\ref{KNNUTAdSmetric}). The field equation for the charged scalar probe is given by
\begin{equation}
(\nabla_{\alpha} - i q A_{\alpha})(\nabla^{\alpha} - i q A^{\alpha})
\Phi = 0\label{KG1},
\end{equation}
where $q$ is the charge of the scalar probe. As we notice, there are two Killing vectors, i.e. $\partial_t$ and $\partial_\phi$ for the Kerr-Newman-NUT-AdS black holes (\ref{KNNUTAdSmetric}), and so, we make the separation of the coordinates in the scalar field, as
\begin{equation}
\Phi(t, r, \theta, \phi) = \mathrm{e}^{- i \omega t + i m \phi} R(r) S(\theta)\label{phi-expand1},
\end{equation}
where $R(r)$ and $S(\theta)$ are two independent functions.
After substituting equation (\ref{phi-expand1}) in (\ref{KG1}), we find two separated differential equations for the angular $S(\theta)$ and the radial $R(r)$ functions \cite{Sakti2018},
\be
\frac1{\sin\theta} \partial_\theta (\sin\theta \, \partial_\theta )S(\theta) - \f{\Xi^2}{\D_\th}\left[ \frac{m^2}{\sin^2\theta}+\f{\left[2n(\cos\theta-1)-a \sin^2\theta\right]^2\omega^2}{\sin^2\theta} \right] S(\theta) \nonumber\\ \ee
\be - \f{\Xi^2}{\D_\th}\left[ 2am\omega \cos\th (4n+a\cos\th) + K_{l'} \right]S(\theta) =0  \label{angular1}, \ee
\be \partial_r (\Delta_r \partial_r)R(r) + \left[ \frac{[ (r^2 + (a+n)^2) \omega - qe  r - m a ]^2\Xi^2}{\Delta_r} + 2 a m \omega\Xi^2 - K_{l'} \right] R(r) = 0\label{radial1}, \ee
where $K_{l'}$ is the separation constant. \textcolor{black}{To simplify the radial and angular equations (\ref{angular1}) and  (\ref{radial1}), and the subsequent calculations, we set the black hole magnetic charge $ g=0$. All our results in this article, can be extended for a black hole where $g \neq 0$, though the results are more complicated and lengthier. }

To simplify the radial equation (\ref{radial1}), we consider a few approximations. First, we consider the near region, which means $ \omega r <<1 $. Second, we assume the low-frequency limit for the probe, which means  $ \omega M <<1 $. Third, we consider the small probe charge, according to $ q e <<1 $. Accordingly, we also impose $ \omega a <<1, \omega e <<1, $ and $ \omega n <<1 $. As we notice from the metric functions (\ref{deltainkerr}), $\Delta_r $ is a quartic polynomial of $r$. Applying the mentioned approximations on $\Delta_r $, we can approximate the quartic polynomial as a quadratic polynomial such as 
\be
\Delta_r \simeq K(r-r_+)(r-r_*), 
\ee 
where $ r_+ $ is the outer horizon of the black hole, and 
\begin{eqnarray} 
K &=& 1\textcolor{black}{+} \f{6r_+^2 +6n^2 +a^2}{l^2}, \\
r_* &=& r_+ - \f{1}{K r_+}\left[r_+^2 -e^2 \textcolor{black}{-} \f{r_+^2(r_+^2 +6n^2 +a^2)}{l^2}\textcolor{black}{-}\f{(3n^2-l^2)(a^2-n^2)}{l^2} \right]. \
\end{eqnarray}
We notice that similar approximation has been used for the quartic metric functions of the Kerr-Newman-AdS black holes \cite{ChenJHEP2010, ChenNuc2011-a} as well as, the Kerr-NUT-AdS black hole \cite{ChenNuc2011-b}. 
We also notice that the Hawking temperature (\ref{en2}) could be approximated as
\begin{equation}
T_H = \f{K(r_+ -r_*)}{4\pi\left[r_+^2 +(a+n)^2\right]\Xi}.
\end{equation}

Applying the above-mentioned approximations to the radial equation (\ref{radial1}), we find 
\begin{eqnarray}
\partial_r (\Delta_r \partial_r)R(r) &+& \left[ \frac{ \Xi^2\left[ (r_+^2 \!+\! (a+n)^2) \omega - a m - qe r_+ \right]^2}{K(r - r_+) (r_+ - r_*)} \!-\! \frac{\Xi^2\left[ (r_*^2 \!+\! (a+n)^2) \omega - a m -qe r_* \right]^2}{K(r - r_*) (r_+ - r_*)} \right] R(r) \nonumber\\
&+&q^2e^2 R(r)= \textcolor{black}{l'(l'+1)} R(r), \label{radialequationapprox}
\end{eqnarray}
where we set the separation constant as $K_{l'}=\textcolor{black}{l'(l'+1)}$. 

The radial differential equation (\ref{radialequationapprox}) in the proper limits, reveals that there are two different individual CFTs that are holographically dual to the Kerr-Newman black hole \cite{othe2,KSChen}. 
The first CFT (which is also known as $J$ picture) is holographically dual to the black hole where the charge of probe is very small. The second CFT (which is also known as $Q$ picture)is holographically dual to the black hole where the probe co-rotates with the horizon. In other words, in the $J$ picture, we set the electric charge of probe to be zero, and in the $Q$ picture, we consider a probe field in the $m = 0$ mode. 
We should note that not necessarily any charged rotating black hole is twofold dual to CFTs. As an counterexample, 
the four-dimensional charged rotating Kerr-Sen black hole is not holographically dual to CFT in the $Q$ picture \cite{Ghezelbash}. Inspired with the existence of two dual CFTs for the Kerr-Newman black holes, in this article, we search for possible dual CFTs to the Kerr-Newman-NUT-AdS black holes \footnote{
Another alternative way to realize two possible dual CFTs to a four-dimensional charged rotating black hole, is to uplift the black hole to five dimensions by adding an internal direction $\chi$ to four dimensions, such that
\be
\Phi (t,r,\theta,\phi,\chi) = \mathrm{e}^{-i\omega t + im\phi + i\textcolor{black}{q}\chi} R(r) S(\theta).\label{5dwave}
\ee
We notice that the internal space $\chi$ leads to a $U(1)$ symmetry along the coordinate $\phi$. 
This extension has been used to investigate the RN/CFT correspondence \cite{RN-CFT,RN-CFT1,RN-CFT2,RN-CFT3,RN-CFT4}. The existence of two coordinates $\phi$ and $\chi$ with $U(1)$ symmetry, provides the twofold hidden symmetry for the charged rotating black holes in $J$ and $Q$ pictures, respectively \cite{Chen-2fold, Ghezelbash,Ghezelbashdeform}.  We also note that an $SL(2,\mathbb{Z})$ modular group transformation for the torus $(\phi,\chi)$ provides merging the two different $J$ and $Q$ pictures into the general picture.}.

\textcolor{black}{We should notice that in the near region and the low-frequency limit, the angular equation (\ref{angular1}) is just an standard Laplacian on the deformed 2-sphere by the NUT charge.  In \cite{Balasubramanian1999}, the authors showed that the symmetry of this equation is $SU(2)\times SU(2) $. So, the angular equation reveals a hidden $SU(2)\times SU(2) $ symmetry, but not $SL(2,\mathbb{R})_L \times SL(2,\mathbb{R})_R$ that we need to describe the hidden conformal symmetry for the Kerr-Newman-NUT-AdS black holes. 
}
\par
\textcolor{black}{In this regard, in section \ref{deformed-hiddensym}, we find a set of conformal generators which produce exactly the deformed radial equation, in different conformal pictures. These conformal generators make an $SL(2,\mathbb{R})_L \times SL(2,\mathbb{R})_R$ symmetry for the deformed radial equation, which also is the symmetry group of AdS$ _3 $ space \cite{Balasubramanian1999}.}
\textcolor{black}{
We should emphasize that the hidden conformal symmetry for the generic Kerr black holes could also be obtained by considering higher spin fields \cite{Lowe2014}.  For different massless spin $ s $ fields, one can find the CFT temperatures and the central charges for the dual CFT to the Kerr black holes. However, we should note the results of \cite{Lowe2014} are based on the separability of the higher spin $s$ wave equations in the background of the Kerr black holes \cite{Cha}. To our knowledge, for the other rotating black holes, such as Kerr-Newman-NUT-AdS, the separability of the higher spin $s$ wave equations have not been explored completely. We leave this open question for a forthcoming article.}

However, In addition, the hidden conformal symmetry of generic Kerr black hole for higher spin field and in particular spin 2 field is found in \cite{Lowe2014}. Therein we can see that the similar CFT temperatures are reproduced for any massless spin $ s $ field around Kerr black hole. Ref. [57] affirmatively support the hidden conformal symmetry. A relation of a scalar field to the CFT is also investigated in \cite{Padmanabhan2002}.

\section{The dual deformed hidden conformal symmetry for the Kerr-Newman-NUT-AdS black holes}\label{deformed-hiddensym}
\subsection{The deformed conformal symmetry in $ J $ picture}

As we notice,  the radial wave equation (\ref{radialequationapprox}) has two poles on $r_+$ and $r_*$, where the function $\Delta_r$ vanishes. 
For the generic Kerr-Newman-NUT-AdS black holes which is far from extremality, we note that $r$ is far enough from $r_*$. Therefore, we can drop the linear and quadratic terms in frequency that arise from expansion of the radial wave equation (\ref{radialequationapprox}), near $r_*$ horizon \cite{Ghezelbash,Lowe,Ghezelbashdeform}. \textcolor{black}{We deform the radial wave equation (\ref{radialequationapprox}) near the horizon $r_*$ by a deformation parameter $\kappa$ such that $ r_* \rightarrow \kappa r_+ $ .} In this section and the next section, we search for the existence of possible $J$ and $Q$ pictures for the Kerr-Newman-NUT-AdS black holes. More specifically, in this section, we consider the deformed radial wave equation in the $J$ picture that can be obtained from (\ref{radialequationapprox}), by setting $q=0$ and is given by
\begin{eqnarray}
\textcolor{black}{\partial_r \left(\Delta_r \partial_r\right)R(r) + K\left[ \f{r_+ - r_*}{r - r_+} A  \!+\! \f{r_+ - r_*}{r - r_*} B + C \right] R(r) = 0,} \label{deformedKNNUTAdS-J}
\end{eqnarray}  \
where
\begin{equation}
 A= \frac{ \Xi^2\left[ (r_+^2 \!+\! (a+n)^2) \omega - a m \right]^2}{K^2 (r_+ - r_*)^2},  \\
\end{equation}
\begin{equation}
B= \frac{\Xi^2\left[ (\kappa^2 r_+^2 \!+\! (a+n)^2) \omega - a m \right]^2}{K^2(r_+ - r_*)^2},~~~ C=- \f{K_{l'}}{K}. \
\end{equation}
The deformation parameter $ \kappa $ should satisfy the following conditions
\begin{equation}
 [\kappa^2 r_+^2 +(a+n)^2]^2 \omega^2 << 2(r_+ -r_*)(r-r_*), \label{ConJ1}
 \end{equation}
\begin{equation}
[\kappa^2 r_+^2 +(a+n)^2]^2 am\omega << 2(r_+ -r_*)(r-r_*), \label{ConJ2}
\end{equation}
to assure that the deformed scalar wave equation (\ref{deformedKNNUTAdS-J}) is still within the low-frequency limit. One feature of the deformation of $ r_* $ is that it doesn't alter the location of other singularities of the radial equation which are located on the outer horizon $ r_+ $ and also on far infinity. 

We consider the following set of vector fields, to find out the deformed hidden conformal symmetry in the $J$ picture, 
\begin{eqnarray}%
L_ \pm &=& \mathrm{e}^{ \pm \rho t \pm \sigma {\phi} } \left( { \mp \sqrt \Delta \partial _r + \frac{{C_1 - \gamma r}}
{{\sqrt \Delta }}\partial _t + \frac{{C_2 - \delta r}}
{{\sqrt \Delta }}\partial _{\phi} } \right) \label{Lpm},\\
L_0 &=& \f{\gamma}{\sqrt{K}} \partial _t + \f{\delta}{\sqrt{K}} \partial _{\phi} , \label{L0} 
\end{eqnarray}
where $\rho,\sigma,\gamma,\delta,C_1$ and $C_2$ are constants. The vector fields $L_\pm$ and $L_0$ form an $SL(2,\mathbb{R})$ algebra  \cite{Ghezelbash,Lowe,Ghezelbashdeform}
\begin{equation}
\left[{L_ \pm ,L_0 } \right] = \pm L_ \pm,~~~\left[ {L_ + ,L_ - } \right] = 2L_0, \label{SL2}
\end{equation}
which is invariant under the automorphisms $ L_\pm\to -L_\pm, ~L_0 \to -L_0 $. We require that the quadratic Casimir operator of $SL(2,\mathbb{R})$ algebra, represents the deformed radial wave equation (\ref{deformedKNNUTAdS-J}). We then find 
\begin{equation}
\mathcal{H}^2=L_0 ^2 - \frac{1}{2}\left( L_+ L_- + L_- L_+\right) = \partial_r \left(\Delta_r \partial_r\right) + K\left[ \f{r_+ - r_*}{r - r_+} A  \!+\! \f{r_+ - r_*}{r - r_*} B \right]. \label{casdef}
\end{equation}
\textcolor{black}{Furnished by the explicit expression (\ref{casdef}) for the quadratic Casimir operator, we can re-derive the deformed scalar radial equation  (\ref{deformedKNNUTAdS-J}), directly from the quadratic Casimir operator as
\begin{equation}
\mathcal{H}^2 R(r)=K_{l'} R(r).
\end{equation}
in the near region and the low-frequency limit. We also note that considering any other coordinate-dependent-coefficients for $\partial_r$ and the Killing vectors $\partial_t,\, \partial _\phi$ in (\ref{Lpm}) and (\ref{L0}), leads to inconsistencies with satisfying the $SL(2,R)$ algebra (\ref{SL2}), as well as with matching  the Casimir operator (\ref{casdef}) to the deformed radial equation (\ref{deformedKNNUTAdS-J}).}

We can determine the constants $\rho,\sigma,\gamma,\delta,C_1$ and $C_2$ from equations (\ref{Lpm}), (\ref{L0}) and (\ref{casdef}).  We find the following two equations for the coefficients of $\partial _r$ in (\ref{casdef}) 
\begin{align}
\rho C_1 + \sigma C_2 +\frac{K}{2}(r_+ + r_*) = 0\label{rhosig1},
\end{align}
and
\begin{align}
K + \rho \gamma + \sigma \delta = 0\label{rhosig2}.
\end{align}
Moreover, the coefficients of $\partial _{\phi} ^2$ and $\partial _t^2$ in (\ref{casdef}) give two other equations as 
\be 
C_2^2 -r_+ r_* \delta ^2 +r\left[(r_+ +r_*)\delta^2 -2C_2\delta \right] = a^2 \Xi^2 ,
\label{cophi2}\ee 
and
\begin{eqnarray}
C_1^2 - r_+ r_* \gamma^2 + r\left[(r_+ + r_*)\gamma^2 -2C_1 \gamma \right] &=& \frac{\Xi^2 \left(r_+^2+(a+n)^2\right)^2(r-r_*)}{r_+ - r_*} \nonumber\\
&-& \frac{\Xi^2 \left(\kappa^2 r_+^2+(a+n)^2\right)^2(r-r_+)}{r_+ - r_*} . \label{t}\
\end{eqnarray}
Finally, we get the following equation, as the coefficient of $\partial _{\phi} \partial _t$ in
(\ref{casdef}) 
\begin{eqnarray}
&-&C_2 C_1 + \gamma \delta \left( {r - r_ + } \right)\left( {r - r_ * } \right) + \delta rC_1 - \delta r^2 \gamma + C_2 \gamma 
\nonumber\\
&=&\frac{2a\Xi^2\left[\left(\kappa^2 r_+^2+(a+n)^2\right)(r-r_+) -\left(r_+^2+(a+n)^2\right)(r-r_*)\right]}{r_+ -r_*} .\label{mt}
\end{eqnarray}
We solve equations (\ref{rhosig1})-(\ref{mt}) for $\rho,\sigma,\gamma,\delta,C_1$ and $C_2$ and find two unique sets of solutions that are given in Table \ref{tab:tab1}. \textcolor{black}{Each set of solutions provide an  $SL(2,\mathbb{R})$ symmetry, and so we get an  $SL(2,\mathbb{R})_L \times SL(2,\mathbb{R})_R$ symmetry based on two branches of the solutions. The  latter symmetry resembles to that of AdS$ _3 $ space \cite{Balasubramanian1999}.}
\begin{center}

\begin{tabular}{ | l | l | l | p{6cm} |}
\hline
& Branch a & Branch b \\ \hline
$\delta$ & $\frac{2 a\Xi}{{r_ + - r_ * }}$ & 0  \\ \hline
$\gamma$ & $\frac{\left[r_+^2\left(\kappa^2 +1 \right)+2(a+n)^2\right]\Xi}{{r_ + - r_ * }}$ & $\frac{r_+^2\left(\kappa^2 -1 \right)\Xi}{{r_ + - r_ * }}$ \\ \hline
$C_1$ & $\frac{\left[(\kappa^2 r_+ +r_*)r_+^2 +(a+n)^2(r_+ +r_*)\right]\Xi}{{r_ + - r_ * }} $ & $\frac{\left[(\kappa^2 r_+ -r_*)r_+^2 +(a+n)^2(r_+ -r_*)\right]\Xi}{{r_ + - r_ * }} $ \\
\hline
$C_2$ & $\frac{a\left(r_+ +r_*\right)\Xi}{{r_ + - r_ * }}$ & $a\Xi$ \\ \hline
$\rho$ & $0$ & $ - \frac{{K(r_ + - r_ * )}}{r_+^2(\kappa^2 -1)\Xi}$ \\ \hline
$\sigma$ & $-\frac{K(r_+ -r_*)}{2a\Xi} $ & $ \frac{K(r_+ -r_*)\left[r_+^2(\kappa^2+1)+2(a+n)^2\right]}{2ar_+^2(\kappa^2-1)\Xi}$ \\ \hline
\end{tabular}

\captionof{table}{Coefficients of conformal generators (\ref{Lpm}) and (\ref{L0}) in the $ J $ picture.} \label{tab:tab1} 
\end{center}

In the limit of $ 1/l^2 =0 $ where $ \Xi =1,K=1 $ and $r_*=r_-$, we find the conformal generators for the deformed dual CFT to the Kerr-Newman-NUT black hole. Furthermore, if we keep $ n $ to be non-zero, but $ e=0 $ \footnote{Note that $ e $ is implicitly in the expressions for the horizons  $ r_+$ and $r_* $.}, we find the conformal generators for the deformed dual CFT to the Kerr-NUT black hole. Moreover, if we consider the limit where $ n =0 $, we find exactly the known conformal generators for the deformed dual CFT to the Kerr-Newman black hole \cite{Ghezelbashdeform}. We note that setting $ e,n,1/l^2=0 $, we find exact results for the deformed dual CFT to the Kerr black hole \cite{Lowe}. 
We find the deformed conformal generators for the Kerr-Newman-AdS, Kerr-NUT-AdS and Kerr-AdS black holes, by setting $ n=0 $, $ e=0 $ and $e=n=0$, respectively in the generators (\ref{Lpm}) and (\ref{L0}). Moreover, substituting $l \rightarrow il$ in equations (\ref{KNNUTAdSmetric})-(\ref{en4}), we find the metric and the relevant physical quantities for the Kerr-Newman-NUT-dS black holes. The same replacement for the cosmological parameter $l$ in the conformal generators (\ref{Lpm}) and (\ref{L0}) and in the Table \ref{tab:tab1}, yields the conformal generators for the dual CFT to the Kerr-Newman-NUT-dS black holes. 

We can write the conformal generators in two branches, explicitly as 
\begin{eqnarray}
L_{\pm}^a &=&
\f{\mathrm{e}^{\mp2\pi T_{R}\phi}}{\sqrt{\Delta}}\left[\frac{\left[r_+^2(\kappa ^2 r_+ +r_*)+(r_+ +r_*)(a+n)^2 -(r_+^2(\kappa^2 +1)+2(a+n)^2)r \right]\Xi}{r_+-r_*}\partial_{t}\right. \nonumber\\
& +&\left. \frac{a(r_+ +r_* -2r)\Xi}{r_+-r_*}\partial_{\phi}\mp\Delta\partial_{r}\right] , \label{F1}\\
L_{0}^a &=& \frac{\left[r_+^2\left(\kappa^2 +1 \right)+2(a+n)^2\right]\Xi}{\sqrt{K} (r_+ - r_ *) }\partial_{t}+\frac{2 a\Xi}{\sqrt{K}{(r_ + - r_ *) }} \partial_{\phi}, \label{LaJ}\
\end{eqnarray}
in branch a, and 
\begin{eqnarray}
{L}_{\pm}^b & = & \f{\mathrm{e}^{\pm\f{K(r_+ -r_*)}{r_+^2(\kappa^2-1)\Xi}t\mp2\pi T_{L}\phi}}{\sqrt{\Delta}}\left[\f{\left[r_+^2 (\kappa^2 r_+ -r_*)+(r_+ -r_*)(a+n)^2 -r_+^2(\kappa^2-1)r\right]\Xi}{r_+ -r_*}\partial_t \right.  \nonumber\\
& +& a\Xi \partial_\phi \mp\Delta\partial_{r}\Bigg{]},\label{F2}\\
{L}_{0}^b& = & \frac{r_+^2\left(\kappa^2 -1 \right)\Xi}{\sqrt{K} (r_+ - r_ *) }\partial_{t}, \label{LbJ} \
\end{eqnarray}
in branch b, respectively. In equations (\ref{F1})-(\ref{LbJ}), $T_R$ and $T_L$ show the temperatures of the CFT, corresponding to the branch a and branch b, respectively, which
are given by
\be
T_R = \frac{K(r_+ -r_*)}{4\pi a\Xi},~~~ T_L = T_R\frac{r_+^2(\kappa^2+1)+2(a+n)^2}{r_+^2(1-\kappa^2)}\label{TLRJ}.
\ee

{\textcolor{black}{The identification of $T_R$ and $T_L$ as the CFT temperatures, is a result of the spontaneously broken partition function of $SL(2,\mathbb{R})_L \times SL(2,\mathbb{R})_R$ theory, to the partition function of $U(1)_L \times U(1)_R$ CFT \cite{stro}, which is given by
\begin{equation}
{\cal Z}= e^{-4\pi^2 T_R L_0^a-4\pi^2 T_L L_0^b}. \label{eq:partitionfunction} 
\end{equation}  } 
{\textcolor{black}{
We note that the vector fields $ L_\pm $ and $L_0,$ are not periodic under the identification $ \phi \sim \phi +2\pi $ of the coordinate $\phi$. By periodic identification of the coordinate $\phi$, the $SL(2,\mathbb{R})_L \times SL(2,\mathbb{R})_R$ symmetry breaks down to a 
$U(1)_L \times U(1)_R$ symmetry with the partition function (\ref{eq:partitionfunction}) \cite{stro,KerrCFT461, Castro}. We also notice that the CFT temperatures $T_R$ and $T_L$, are independent of the Hawking temperature (\ref{en2}). We note that in the limits of $n=0$ and $\Lambda=0$, the conformal generators (\ref{F1}), (\ref{LaJ}), (\ref{F2}) and (\ref{LbJ}) reduce exactly,  to the dual hidden conformal generators for the Kerr-Newman black holes, in the $J$ picture \cite{Ghezelbashdeform}. }
{So far, we have concluded there are some signatures for existence of the dual CFT to the Kerr-Newman-NUT-AdS black hole. One crucial feature of any CFT is the central charge of the conformal algebra. Though there is an explicit way to calculate the central charge of a CFT dual to the extremal black holes \cite{stro}, however the formalism can't be applied to the non-extremal black holes.  For an extremal black hole, there is only a left sector in the dual CFT with corresponding non-zero left temperature. For a non-extremal black hole, both right and left sectors exist in the dual CFT, with two non-zero left and right temperatures. We assume that the central charges of the dual CFT to the non-extremal black holes should smoothly approach to those of dual CFT to the extremal black holes, in the extremality limit. So, we propose the following left- and right-moving central charges for the dual CFT in the $ J$ picture, as
\be
c_L = c_R = \frac{6a r_+^2(1 -\kappa^2)}{K(r_+ -r_*)}.\label{cJ}
\ee

The microscopic entropy for the dual CFT with two right and left sectors, is given by the Cardy entropy
\be
S_{Cardy} = \frac{\pi^2}{3}(c_R T_R + c_L T_L),\label{Sgen}
\ee
where the central charges and the temperatures are given by (\ref{cJ}) and (\ref{TLRJ}), respectively. We find that the Cardy entropy for the dual CFT to the Kerr-Newman-NUT-AdS black holes, is 
\begin{equation}
S_{Cardy} =\f{\pi}{\Xi} \left[r_+^2+(a+n)^2\right]. \label{CardyBHentropy} \
\end{equation}
The Bekenstein-Hawking entropy for the Kerr-Newman-NUT-AdS black holes is given by (\ref{en1}), which leads to
\be
S_{Cardy} =S_{BH} .
\ee
}
\textcolor{black}{We also should mention about the reproduction of the Cardy formula, from the near horizon geometry of the black holes, that lead to the Virasoro algebra \cite{CarlipGREG2007,CarlipIJMPD2008,Majhi2012,Chakraborty2016}. As we notice, the temperatures (\ref{TLRJ}) and the central charges (\ref{cJ}) of the CFTs, depend non-trivially on the deformation parameter $\kappa$.   We note that we did not derive the central charges (\ref{cJ}) from the asymptotic symmetry group (ASG) or the stretched horizon methods. We just propose them, to match the Cardy entropy of the CFT to the Bekenstein-Hawking entropy of the black holes.  However, it would be very interesting to derive the central charges (\ref{cJ}) of the deformed CFTs using either ASG or stretched horizon techniques \cite{stro,Compere,RN-CFT,RN-CFT1,RN-CFT2,RN-CFT3}. } \textcolor{black}{Furthermore, it also would be fascinating to relate the Cardy and Bekenstein-Hawking entropies with the entanglement entropy as what has been done by Azeyanagi \textit{et. al.} \cite{AzeyanagiTakayanagiPRD}. They compute the holographic entanglement entropy via AdS/CFT corespondence for the BTZ black hole in $ (2+1) $ dimensions. It is shown that this entanglement entropy is equal to Bekenstein-Hawking entropy with the entanglement correction term in which the Bekesnstein-Hawking entropy matches the entropy from CFT \footnote{\textcolor{black}{The authors in \cite{RyuTakayanagiPRD,RyuTakayanagiJHEP}, compute the entanglement entropy from the holographic prescription, that is given by the area of the minimal surface at constant time,
\begin{equation}
S_A = \frac{\text{Area}(\gamma_A)}{4G_N^{(d+2)}}, \nonumber\
\end{equation}
where $ \gamma_A $ is the (unique) minimal surface in AdS$ _{d+2} $ whose boundary coincides with the boundary of the region $ A $. The setup of this calculation is the AdS$ _{d+2} $ space with the Newton constant $ G_N^{(d+2)} $ which is dual to a CFT$ _{d+1} $. In that work, they compute for low and high temperature assumption. In high temperature calculation, it is shown that the entanglement entropy of the black hole is the Bekenstein-Hawking entropy plus the correction term. To confirm that computation, they also calculate the entropy using Cardy formula from AdS$ _3 $/CFT$ _2 $ correspondence since the generic BTZ black hole's degree of freedom is dual to the CFT degree of freedom. They finally prove that both calculations are in agreement, so the entanglement entropy of the black hole is also the entropy from CFT with the correction.}}. We leave studying the entanglemnet entropy in the background of the Kerr-Newman-NUT-AdS k holes for future.}

\subsection{The deformed conformal symmetry in $ Q $ picture}

In $Q$ picture, we set the angular momentum of the probe field $ m =0$ and keep $ q $ to be a non-zero constant. We then find that the deformed radial wave equation in $Q$ picture, is given by
\begin{equation}
\partial_r \left[(r-r_+)(r-r_*) \partial_r\right]R(r) + \left[ \f{r_+ - r_*}{r - r_+} A  \!+\! \f{r_+ - r_*}{r - r_*} B + C \right] R(r) = 0,
\label{deformedKNNUTAdS-QQ} 
\end{equation}
where the constants $A,B$ and $C$ are given by
\begin{equation}
 A= \frac{ \Xi^2\left[ (r_+^2 \!+\! (a+n)^2) \omega - eqr_+ \right]^2}{K^2 (r_+ - r_*)^2}, \\
\end{equation}
\begin{equation}
B= \frac{\Xi^2\left[ (\kappa^2 r_+^2 \!+\! (a+n)^2) \omega - eq\kappa r_+ \right]^2}{K^2(r_+ - r_*)^2},~~~ C=\f{e^2q^2 - K_{l'}}{K}, \
\end{equation}
respectively. 
In $ Q $ picture, we require that the deformation parameter $ \kappa $ satisfies the following conditions
\begin{equation}
 [\kappa^2 r_+^2 +(a+n)^2]^2 \omega^2 << 2(r_+ -r_*)(r-r_*), \label{conQ1}
\end{equation}
\begin{equation}
[\kappa^2 r_+^2 +(a+n)^2]^2 e\kappa q\omega << 2(r_+ -r_*)(r-r_*), \label{conQ2}
\end{equation}
to make sure that the deformed scalar wave equation (\ref{deformedKNNUTAdS-QQ}) is within the low-frequency limit. We note that the first condition (\ref{conQ1}) is the same as (\ref{ConJ1}), while the second condition (\ref{conQ2}) is quite different than (\ref{ConJ2}). Similar to $ J $ picture, to reveal the deformed hidden conformal symmetry, we consider the set of vector fields (\ref{Lpm}) and (\ref{L0}). The vector fields $L_\pm$ and $L_0$ form the $SL(2,\mathbb{R})$ algebra with constants $\rho,\sigma,\gamma,\delta,C_1, C_2$. The two equations for the $\partial _r$,  are the same as equations (\ref{rhosig1}) and (\ref{rhosig2}), while the equation for the  $\partial _t^2$ is the same as (\ref{t}). Moreover, the coefficient of $\partial _{\chi} ^2$  gives the following equation
\be 
C_2^2 -r_+ r_* \delta ^2 +r\left[(r_+ +r_*)\delta^2 -2C_2\delta \right] = \f{ \Xi^2e^2 r_+^2 \left[(r-r_*) - \kappa^2(r-r_+)\right]}{r_+ -r_*} .
\label{cochi2}\ee 
We also find the following equation for the coefficient of $\partial _{\chi} \partial _t$, as  
\begin{eqnarray}
&-& C_2 C_1 + \gamma \delta \left( {r - r_ + } \right)\left( {r - r_ * } \right) + \delta rC_1 - \delta r^2 \gamma + C_2 \gamma 
\nonumber\\
&=&\frac{2er_+\Xi^2\left[\left(\kappa^2 r_+^2+(a+n)^2\right)\kappa(r-r_+) -\left(r_+^2+(a+n)^2\right)(r-r_*)\right]}{r_+ -r_*} .\label{chit}
\end{eqnarray}
The two classes of solutions to the equations (\ref{rhosig1}), (\ref{rhosig2}), (\ref{t}), (\ref{cochi2})  and (\ref{chit}), are given in Table \ref{tab:tab2}.
\begin{center}

\begin{tabular}{ | l | l | l | p{6cm} |}
\hline
& Branch a & Branch b \\ \hline
$\delta$ & $\frac{er_+(\kappa +1)\Xi}{{r_ + - r_ * }}$ & $\frac{er_+(\kappa -1)\Xi}{{r_ + - r_ * }}$  \\ \hline
$\gamma$ & $\frac{\left[r_+^2\left(\kappa^2 +1 \right)+2(a+n)^2\right]\Xi}{{r_ + - r_ * }}$ & $\frac{r_+^2\left(\kappa^2-1\right)\Xi}{{r_ + - r_ * }}$ \\ \hline
$C_1$ & $\frac{\left[(\kappa^2 r_+ +r_*)r_+^2 +(a+n)^2(r_+ +r_*)\right]\Xi}{{r_ + - r_ * }} $ & $\frac{\left[(\kappa^2 r_+ -r_*)r_+^2 +(a+n)^2(r_+ -r_*)\right]\Xi}{{r_ + - r_ * }} $ \\
\hline
$C_2$ & $\frac{er_+(\kappa r_+ +r_*)\Xi}{{r_ + - r_ * }}$ & $\frac{er_+(\kappa r_+ -r_*)\Xi}{{r_ + - r_ * }}$ \\ \hline
$\rho$ & $\frac{K(r_+ -r_*)}{2\left(\kappa r_+^2 - (a+n)^2\right)\Xi}$ & $ \frac{{K(r_ + - r_ * )(\kappa -1)}}{2(1-\kappa)\left(\kappa r_+^2 - (a+n)^2\right)\Xi}$ \\ \hline
$\sigma$ & $\frac{Kr_+(r_+ -r_*)(\kappa +1)}{2e\left(\kappa r_+^2 -(a+n)^2\right)\Xi} $ & $ \frac{K(r_+ -r_*)\left[r_+^2(\kappa^2+1)+2(a+n)^2\right]}{2er_+(\kappa -1)\left(\kappa r_+^2 - (a+n)^2\right)\Xi}$ \\ \hline
\end{tabular}
\captionof{table}{Coefficients of conformal generators (\ref{Lpm}) and (\ref{L0}) in the $ Q $ picture.} \label{tab:tab2} 
\end{center}

The explicit expressions for the first and second sets of the deformed conformal generators in $Q$ picture, are given by
\begin{eqnarray}
L_{\pm}^a &=&
\f{\mathrm{e}^{\pm\frac{K(r_+ -r_*)}{2\left(\kappa r_+^2 - (a+n)^2\right)\Xi} t \mp 2\pi T_{R}\phi}}{\sqrt{\Delta}}\left[\mp\Delta\partial_{r}+\frac{er_+\left((\kappa r_+ +r_*)-(\kappa +1)r\right)\Xi}{r_+-r_*}\partial_{\chi}\right. \nonumber\\
& +&\left. \frac{\left[r_+^2(\kappa ^2 r_+ +r_*)+(r_+ +r_*)(a+n)^2 -(r_+^2(\kappa^2 +1)+2(a+n)^2)r \right]\Xi}{r_+-r_*}\partial_{t}\right] , \label{F3}\\
L_{0}^a &=& \frac{\left[r_+^2\left(\kappa^2 +1 \right)+2(a+n)^2\right]\Xi}{\sqrt{K} (r_+ - r_ *) }\partial_{t}+\frac{er_+(\kappa +1)\Xi}{\sqrt{K}{(r_ + - r_ *) }} \partial_{\chi}, \label{LaQ}\
\end{eqnarray}
for branch a 
\begin{eqnarray}
{L}_{\pm}^b & = & \f{\mathrm{e}^{\pm\f{K(r_+ -r_*)(\kappa+1)}{2(1-\kappa)\left(\kappa r_+^2 - (a+n)^2\right)\Xi}t\mp2\pi T_{L}\phi}}{\sqrt{\Delta}}\left[\f{\left[r_+^2 (\kappa^2 r_+ -r_*)+(r_+ -r_*)(a+n)^2 -r_+^2(\kappa^2-1)r\right]\Xi}{r_+ -r_*}\partial_t \right.  \nonumber\\
&+ & \f{er_+\left[(\kappa r_+-r_*)-(\kappa-1)r\right]\Xi}{r_+ -r_*} \partial_\chi \mp\Delta\partial_{r}\Bigg{]},\label{F4}\\
{L}_{0}^b& = & \frac{r_+^2\left(\kappa^2 -1 \right)\Xi}{\sqrt{K} (r_+ - r_ *) }\partial_{t} + \f{er_+(\kappa-1)\Xi}{\sqrt{K}(r_+ -r_*)}\partial_\chi, \label{LbQ} \
\end{eqnarray}
and for branch b, respectively.

\textcolor{black}{Similar to $ J $ picture, the two sets of conformal generators represent the hidden symmetry  $SL(2,\mathbb{R})_L \times SL(2,\mathbb{R})_R$ of the radial wave equation. We find that the deformed radial equation (\ref{deformedKNNUTAdS-QQ}) can be re-written as 
\begin{equation}
\mathcal{H}^2 R(r)=K_{l'} R(r).\
\end{equation}
where $\mathcal{H}^2 $ is the Casimir operator of the conformal algebra, constructed from the generators (\ref{F3})-(\ref{LaQ}) or (\ref{F4})-(\ref{LbQ}).}
The finite right- and left-moving temperatures in (\ref{F3})-(\ref{LbQ}),  are given by
\be
T_R = \frac{Kr_+(r_+ -r_*)(\kappa +1)}{4\pi e\left(\kappa r_+^2 -(a+n)^2\right)\Xi},~~~ T_L = T_R\frac{r_+^2(\kappa^2+1)+2(a+n)^2}{r_+^2(1-\kappa^2)}\label{TLRQ},
\ee
respectively. We also conjecture the central charges as
\be
c_L = c_R = \frac{6e r_+(1 -\kappa)\left(\kappa r_+^2-(a+n)^2\right)}{K(r_+ -r_*)}.\label{cQ}
\ee
The CFT temperatures (\ref{TLRQ}) and the central charges (\ref{cQ}), lead to the Cardy entropy
\be
S_{Cardy} = \frac{\pi^2}{3}(c_R T_R + c_L T_L)=\f{\pi}{\Xi} \left[r_+^2+(a+n)^2\right], \label{CardyBHentropyQ} 
\ee
which is exactly in agreement with the Bekenstein-Hawking entropy (\ref{en1}) for the 
Kerr-Newman-NUT-AdS black holes.
We also note that in the limits of $n=0$ and $1/l^2=0$, the conformal generators (\ref{F3})-(\ref{LbQ}) reduce exactly,  to the dual deformed hidden conformal generators for the Kerr-Newman black holes in $Q$ picture \cite{Ghezelbashdeform}. Moreover, in the limit of $ 1/l^2 =0 $ where $ \Xi =1,K=1 $ and $r_*=r_-$, we find the conformal generators for the deformed dual CFT to the Kerr-Newman-NUT black hole. Furthermore, if we keep $ n $ to be non-zero, but $ e=0 $, we find the conformal generators for the deformed dual CFT to the Kerr-NUT black hole. Moreover, if we consider the limit where $ n =0 $, we find exactly the known conformal generators for the deformed dual CFT to the Kerr-Newman black hole \cite{Ghezelbashdeform}. We note that setting $ e,n,1/l^2=0 $, we find exact results for the deformed dual CFT to the Kerr black hole \cite{Lowe}. 
We also find the deformed conformal generators for the Kerr-Newman-AdS, Kerr-NUT-AdS and Kerr-AdS black holes, by setting $ n=0 $, $ e=0 $ and $e=n=0$, respectively in the generators (\ref{Lpm}) and (\ref{L0}) .

\subsection{The dual deformed conformal symmetry in general picture}
In this section, we find the deformed conformal generators for the dual CFT,  to the Kerr-Newman-NUT-AdS black holes, in the general picture. The modular group $SL(2,\mathbb{Z})$ consists of the following transformation on the torus $(\phi,\chi)$ \cite{Ghezelbash,chen-novelsl2z,Chen-genhidden4d5d,Ghezelbashdeform}.
\begin{equation}
\left( {\begin{array}{*{20}c}
{\phi '} \\
{\chi '} \\
\end{array}} \right) = \left( {\begin{array}{*{20}c}
\alpha & \beta \\
\eta & \tau \\
\end{array}} \right)\left( {\begin{array}{*{20}c}
\phi \\
\chi \\
\end{array}} \right),\label{SL2trans}
\end{equation} 
where $
\left( {\begin{array}{*{20}c}
\alpha & \beta \\
\eta & \tau \\
\end{array} } \right) 
$
is the $SL(2,\mathbb{Z})$ group element and the coordinates $\phi '$ and $\chi '$ are the general angular coordinates \cite{chen-novelsl2z}. In general picture, the phase factor of the charged scalar field (\ref{5dwave}), is invariant under modular transformation (\ref{SL2trans}). So, we find that
\begin{equation}
m = \alpha m' + \eta q'~,~q = \beta m' + \tau q'.\label{em-em}
\end{equation}
In the general picture, we have two possible options. First, we can set $q' = 0$ and keep $m'\ne 0$, which we call the $J'$ picture. Second,  we can set $m' = 0$ and keep $q'\ne 0$, which we call the $Q'$ picture. In the first picture, the deformed radial wave equation becomes
\begin{equation}
\partial_r \left[(r-r_+)(r-r_*) \partial_r\right]R(r) + \left[ \f{r_+ - r_*}{r - r_+} A  \!+\! \f{r_+ - r_*}{r - r_*} B + C \right] R(r) = 0,\label{dXXX} \
\end{equation} 
where
\begin{equation}
 A= \frac{ \Xi^2\left[ (r_+^2 \!+\! (a+n)^2) \omega - a_1 m' \right]^2}{K^2 (r_+ - r_*)^2}, 
\end{equation}
\begin{equation}
B= \frac{\Xi^2\left[ (\kappa^2 r_+^2 \!+\! (a+n)^2) \omega - a_2 m' \right]^2}{K^2(r_+ - r_*)^2},~~~ C= \f{e^2 q^2-K_{l'}}{K}, \
\end{equation}
and
\begin{equation}
a_1 = a\alpha +e\beta r_+,~~~ a_2 =  a\alpha +e\beta \kappa
r_+.
\end{equation}
The deformation parameter $ \kappa $, in $J'$ picture, should satisfy the following conditions
\begin{equation}
 \left[\kappa^2 r_+^2 +(a+n)^2\right]^2 \omega^2 << 2(r_+ -r_*)(r-r_*), \
\end{equation}
\begin{equation}
\left[\kappa^2 r_+^2 +(a+n)^2\right]^2 a_2 m'\omega << 2(r_+ -r_*)(r-r_*), \
\end{equation}
to make sure that the deformed scalar wave equation, is within the low-frequency limit. 
After constructing the quadratic Casimir operator of $SL(2,\mathbb{R})$ algebra, that should match to the deformed radial wave equation (\ref{dXXX}), we find the two equations for the $\partial _r$,  are the same as equations (\ref{rhosig1}) and (\ref{rhosig2}), while the equation for the  $\partial _t^2$ is the same as (\ref{t}).  However, the equation for the $\partial _{\phi'} ^2$ yields
\be 
C_2^2 -r_+ r_* \delta ^2 +r\left[(r_+ +r_*)\delta^2 -2C_2\delta \right] = \f{ \Xi^2 \left[a_1^2(r-r_*) - a_2^2(r-r_+))\right]}{r_+ -r_*}.
\label{cophi3}\ee 
Moreover, the  equation for the  $\partial _{\phi'} \partial _t$,  is given by
\begin{eqnarray}
& -&C_2 C_1 + \gamma \delta \left( {r - r_ + } \right)\left( {r - r_ * } \right) + \delta rC_1 - \delta r^2 \gamma + C_2 \gamma 
\nonumber\\
&=&\frac{2\Xi^2\left[a_2\left(\kappa^2 r_+^2+(a+n)^2\right)(r-r_+) -a_1\left(r_+^2+(a+n)^2\right)(r-r_*)\right]}{r_+ -r_*} .\label{phit}
\end{eqnarray}
We find two different classes of solutions to the equations (\ref{rhosig1}), (\ref{rhosig2}), (\ref{t}), (\ref{cophi3}) and (\ref{phit}), that are given in Table \ref{tab:tab3}, respectively.
\begin{center}
\begin{tabular}{ | l | l | l | p{6cm} |}
\hline
& Branch a & Branch b \\ \hline
$\delta$ & $\frac{\left(a_1 +a_2 \right)\Xi}{{r_ + - r_ * }}$ & $\frac{\left(a_1 - a_2 \right)\Xi}{{r_ + - r_ * }}$  \\ \hline
$\gamma$ & $\frac{\left[r_+^2\left(\kappa^2 +1 \right)+2(a+n)^2\right]\Xi}{{r_ + - r_ * }}$ & $\frac{r_+^2\left(\kappa^2 -1\right)\Xi}{{r_ + - r_ * }}$ \\ \hline
$C_1$ & $\frac{\left[(\kappa^2 r_+ +r_*)r_+^2 +(a+n)^2(r_+ +r_*)\right]\Xi}{{r_ + - r_ * }} $ & $\frac{\left[(\kappa^2 r_+ -r_*)r_+^2 +(a+n)^2(r_+ -r_*)\right]\Xi}{{r_ + - r_ * }} $ \\
\hline
$C_2$ & $\frac{\left(a_1 r_* + a_2 r_+\right)\Xi}{{r_ + - r_ * }}$  & $\frac{(a_1 r_* - a_2 r_+)\Xi}{{r_ + - r_ * }}$ \\ \hline
$\rho$ & $\frac{-K(a_1 -a_2)(r_+ -r_*)}{2\left[r_+^2(a_1\kappa^2 -a_2 ) + (a+n)^2(a_1-a_2)\right]\Xi}$ & $ \frac{-K(a_1 +a_2)(r_+ -r_*)}{2\left[r_+^2(a_1\kappa^2 -a_2 ) + (a+n)^2(a_1-a_2)\right]\Xi}$ \\ \hline
$\sigma$ & $\frac{Kr_+^2(r_+ -r_*)(1-\kappa^2)}{2\left[r_+^2(a_1\kappa^2 -a_2 ) + (a+n)^2(a_1-a_2)\right]\Xi} $ & $ \frac{K(r_+ -r_*)\left[r_+^2(\kappa^2 +1)+2(a+n)^2 \right]}{2\left[r_+^2(a_1\kappa^2 -a_2 ) + (a+n)^2(a_1-a_2)\right]\Xi}$ \\ \hline
\end{tabular}
\captionof{table}{Coefficients of conformal generators (\ref{Lpm}) and (\ref{L0}) in the $ J' $ picture.} \label{tab:tab3} 
\end{center}
We notice that the results in Table \ref{tab:tab3}, reduce perfectly to those in Table \ref{tab:tab1} for $J$ picture, if we set $\alpha=1$ and $\beta=0$. Moreover, the results in Table \ref{tab:tab3}, reduce to those in Table \ref{tab:tab2} for $Q$ picture, if we set $\alpha=0$ and $\beta=1$.
The explicit expressions for the deformed conformal generators in $J'$ picture, are given by
\begin{eqnarray}
L_{\pm}^a &=&
\f{\mathrm{e}^{\pm\frac{-K(a_1 -a_2)(r_+ -r_*)}{2\left[r_+^2(a_1\kappa^2 -a_2 ) + (a+n)^2(a_1-a_2)\right]\Xi} t \mp 2\pi T_{R}\phi'}}{\sqrt{\Delta}}\left[\mp\Delta\partial_{r}+\frac{a_2(r_+ - r)+a_1(r_* -r)\Xi}{r_+-r_*}\partial_{\phi'}\right. \nonumber\\
&+ &\left. \frac{\left[r_+^2(\kappa ^2 r_+ +r_*)+(r_+ +r_*)(a+n)^2 -(r_+^2(\kappa^2 +1)+2(a+n)^2)r \right]\Xi}{r_+-r_*}\partial_{t}\right], \label{Lpm1} \\
L_{0}^a &=& \frac{\left[r_+^2\left(\kappa^2 +1 \right)+2(a+n)^2\right]\Xi}{\sqrt{K} (r_+ - r_ *) }\partial_{t}+\frac{(a_1 +a_2)\Xi}{\sqrt{K}{(r_ + - r_ *) }} \partial_{\phi'}, \label{LaG}\
\end{eqnarray}
for branch a and  
\begin{eqnarray}
L_{\pm}^b &=&
\f{\mathrm{e}^{\pm\frac{-K(a_1 +a_2)(r_+ -r_*)}{2\left[r_+^2(a_1\kappa^2 -a_2 ) + (a+n)^2(a_1-a_2)\right]\Xi} t \mp 2\pi T_{L}\phi'}}{\sqrt{\Delta}}\left[\mp\Delta\partial_{r}+\frac{a_2(r_+ - r)-a_1(r_* -r)\Xi}{r_+-r_*}\partial_{\phi'}\right. \nonumber\\
& +&\left. \frac{\left[r_+^2(\kappa ^2 r_+ -r_*)+(r_+ -r_*)(a+n)^2 - r_+^2(\kappa^2 -1)r \right]\Xi}{r_+-r_*}\partial_{t}\right] , \label{Lpm2} \\
L_{0}^b &=& \frac{r_+^2\left(\kappa^2 -1 \right)\Xi}{\sqrt{K} (r_+ - r_ *) }\partial_{t}+\frac{(a_2-a_1)\Xi}{\sqrt{K}{(r_ + - r_ *) }} \partial_{\phi'}, \label{LbG}\
\end{eqnarray}
for branch b, respectively, where the finite CFT temperatures are 
\be
T_R = \frac{Kr_+^2(r_+ -r_*)(\kappa^2 -1)}{4\pi\left[r_+^2(a_1\kappa^2 -a_2 ) + (a+n)^2(a_1-a_2)\right]\Xi},~~~ T_L = T_R\frac{r_+^2(\kappa^2+1)+2(a+n)^2}{r_+^2(1-\kappa^2)}\label{TLRG}.
\ee
We note that, so far, the relation between the left-moving temperature $T_L$ versus the right-moving temperature  $T_R$ is the same in all different CFT pictures. 
\textcolor{black}{We notice that similar to $J$ and $Q$ pictures, the two sets of conformal generators (\ref{Lpm1})-(\ref{LbG}), represent $SL(2,\mathbb{R})_L \times SL(2,\mathbb{R})_R$ symmetry. Moreover, we can re-write the radial equation (\ref{dXXX}) as
\begin{equation}
\mathcal{H}^2 R(r)=K_{l'} R(r).\
\end{equation}
}
In $J'$ picture,
we find exact agreement 
between the Bekenstein-Hawking entropy (\ref{en1}) and the Cardy entropy (\ref{Sgen}), if we consider the central charges of the CFT, as
\be
c_L = c_R = \frac{6\left[r_+^2(a_2 -a_1\kappa^2)+(a+n)^2(a_2 -a_1)\right]}{K(r_+ -r_*)}.\label{cG}
\ee
The temperatures (\ref{TLRG}) and the central charges (\ref{cG}) of the CFTs, depend non-trivially on the deformation parameter $\kappa$.   We also note that  we did not derive the central charges (\ref{cG}) from the asymptotic symmetry group (ASG) or the stretched horizon methods. We just propose them, from the matching of the Bekenstein-Hawking entropy and the Cardy entropy.   However, it would be very interesting to derive the central charges (\ref{cG}) of the deformed CFTs using either ASG or stretched horizon techniques \cite{stro,Compere,RN-CFT,RN-CFT1,RN-CFT2,RN-CFT3}. We also note that in the limits of $n=0$ and $\Lambda=0$, the conformal generators (\ref{Lpm1}), (\ref{LaG}), (\ref{Lpm2}) and (\ref{LbG}) reduce exactly,  to the dual deformed hidden conformal generators for the Kerr-Kerr-Newman black holes in $J'$ picture \cite{Ghezelbashdeform}.

The deformed scalar wave equation, in $Q '$ picture, is given by equation (\ref{dXXX}), with following replacements for the modular group parameters: $\alpha \rightarrow \eta$ and $\beta \rightarrow \tau$, as well as replacing the mode number $m' \rightarrow q'$. A similar calculation shows that, we get the temperatures (\ref{TLRG}) and central charges (\ref{cG}) in $Q'$ picture with the above replacements for the modular group parameters and the mode number. Finally, we find two sets of deformed hidden conformal generators for the $Q'$ picture,  given by expressions (\ref{Lpm1})-(\ref{LbG}), with replacements
$\alpha \rightarrow \eta,\, \beta \rightarrow \tau$ and $m' \rightarrow q'$. We have explicitly checked out , in the limits of $n=0$ and $\Lambda=0$, the dual deformed hidden conformal generators reduce exactly, to the dual deformed conformal generators for the Kerr-Kerr-Newman black holes in $Q'$ picture \cite{Ghezelbashdeform}.

Moreover, we find that setting the deformation parameter $\kappa=r_*/r_+$ in the deformed generators (\ref{F1})-(\ref{LbJ}), (\ref{F3})-(\ref{LbQ}) and (\ref{Lpm1})-(\ref{LbG}) yields the generators of the hidden conformal symmetry for the Kerr-Newman-NUT-AdS black holes \cite{Sakti2018}. Similar behaviour was noticed in \cite{Ghezelbashdeform}, where one set $\kappa =r_-/r_+$, in the deformed conformal generators of the Kerr-Newman black holes. 

A very interesting feature of the deformed conformal generators for the Kerr-Newman-NUT-AdS, is that, they provide a realization of the conformal generators for the non-rotating Reissner-Nordstr{\"o}m-NUT-AdS black holes. We discuss about the conformal generators for the Reissner-Nordstr{\"o}m-NUT-AdS black holes in section \ref{deformedRNNUTAdS}.

\section{Scattering of the charged scalar fields, in the background of Kerr-Newman-NUT-AdS black holes}\label{scat}

In this section, we consider the absorption cross-section of the scalar fields in the background of Kerr-Newman-NUT-AdS black holes in general $J '$ picture. We consider the following coordinate transformations from $(t,r,\phi)$ to the near-extremal near-horizon coordinates $(\tau,y,\varphi)$ \cite{ChenJHEP2010, ChenNuc2011-a,ChenNuc2011-b}
\begin{eqnarray}
r = \f{r_+ + r_*}{2}+\lambda r_0 y, ~~r_+ -r_* = \mu_1 \lambda r_0,~~ t = \frac{r_0 \Xi}{\lambda}\tau, ~~ \phi = \varphi + \frac{\Omega_H r_0 \Xi}{\lambda}\tau, \label{nearextremaltransformation} \
\end{eqnarray}
where $r_0=\sqrt{\frac{r_+^2+a^2}{K}}$ and $\lambda \rightarrow 0$ shows the near-horizon limit.  We also note that we should impose
one more condition
\begin{equation}
r_+ -\kappa r_+ = \mu_2 \lambda r_0,
\end{equation}
where for the non-deformed near-extremal near-horizon geometry,  $ \mu_2=\mu_1 $ \cite{Sakti2018}. For the frequencies $\omega$, near the superradiant bound $ \omega_s $
\begin{equation}
\omega = \omega_s +\hat{\omega}\frac{\lambda}{r_0},
\end{equation}
where  $\omega_s = m\Omega_H + q\Phi_H$.
In near-extremal near-horizon geometry, the radial equation becomes
\begin{eqnarray}
\left[\partial_y\left(y-\f{\mu_1}{2}\right)\left(y+\f{\mu_1}{2}\right)\partial_y +\f{A_s}{y-\f{\mu_1}{2}}+\f{B_s}{y+\f{\mu_1}{2}} +C_s\right]R(y) =0,\label{radialequationnearext}
\end{eqnarray}
where
\begin{eqnarray}
A_s = \frac{\hat{\omega}^2\Xi^2}{\mu_1}, ~~~ B_s = -\f{\mu_2^2\Xi^2}{\mu_1}\left(\frac{\hat{\omega}}{\mu_2}-\frac{2 m\Omega_H r_+}{K} +\frac{qe}{K}\f{(a+n)^2-r_+^2}{(a+n)^2+r_+^2} \right)^2,\
\end{eqnarray}
and $ C_s $ is the separation constant in the angular equation. The angular equation is independent of the frequency, but we don't consider it anymore. To solve the differential equation (\ref{radialequationnearext}),  we consider a coordinate transformation $ z =\frac{r-\mu_1/2}{r+\mu_1/2} $, where the the radial equation (\ref{radialequationnearext}) becomes
\begin{eqnarray}
\left[z(1-z)\partial_z^2 + (1-z)\partial_z +\f{\hat{A_s}}{z}+\hat{B_s}+\f{C_s}{1-z}\right]R(z) =0,\label{radialequationnearext0}
\end{eqnarray}
where $
\hat{A_s} = A_s/\mu_1,\hat{B_s} = B_s/\mu_1 $. The solution to (\ref{radialequationnearext0}) are given by
\begin{equation}
R(z)= z^{\alpha}(1-z)^{\beta}F(a_s,b_s,c_s;z), \label{SOL}
\end{equation}
where $ F(a_s,b_s,c_s;z) $ is the hypergeometric function. The parameters of the hypergeometric functions are given by 
\textcolor{black}{
\begin{equation}
a_s = \beta_s +i(\gamma_s-\alpha_s), ~b_s =\beta_s -i(\gamma_s+\alpha_s),~c_s = 1- 2i\alpha_s, \
\end{equation}
where
\begin{equation}
\alpha_s = \sqrt{\hat{A_s}}, ~\beta_s = \frac{1}{2}\left(1-\sqrt{1-4C_s}\right),~\gamma_s = \sqrt{-\hat{B}_s}. \
\end{equation}
}
In the very far limit, where  $ (y>>\mu_1/2 ) \sim (z= 1)$, the solutions (\ref{SOL})  behave as
\begin{eqnarray}
R(y) \sim D_1 y^{-\beta}+D_2 y^{\beta-1},
\end{eqnarray}
where
\begin{equation}
D_1 = \frac{\Gamma(c_s)\Gamma(2h-1)}{\Gamma(c_s-a_s)\Gamma(c_s-b_s)}, ~~~ D_2 = \frac{\Gamma(c_s)\Gamma(1-2h)}{\Gamma(a_s)\Gamma(b_s)}, ~~~h= 1-\beta_s, 
\end{equation}
and $ h $ is the conformal weight. The essential properties of the absorption cross section is captured by the coefficient $ D_1 $ such as
\begin{align}
P_{abs} \sim \left| D_1 \right|^{-2} = \frac{\sinh \left( {2\pi\alpha } \right)}{2\pi \alpha}\frac{{\left| {\Gamma \left( c_s - a_s  \right)} \right|^2 \left| {\Gamma \left( c_s - b_s \right)} \right|^2 }}{{ \left( {\Gamma \left( {2h-1} \right)} \right)^2 }}\label{PabsSL2Z}.
\end{align}
The real-time correlator could be read alternatively as \cite{ChenChu,ChenLong}
\begin{eqnarray}
G_R \sim \frac{D_2}{D_1} = \frac{\Gamma(1-2h)}{\Gamma(2h-1)} \frac{\Gamma(c_s-a_s)\Gamma(c_s-b_s)}{\Gamma(a_s)\Gamma(b_s)} ,
\end{eqnarray}
where $ D_1 $ as the source and $ D_2 $ as the response.

The main objective of the Kerr/CFT correspondence is to find the agreement between the absorption cross-section (\ref{PabsSL2Z}) and the absorption cross-section in a two-dimensional CFT. In this paper, we wish to prove the agreement between the absorption cross-section (\ref{PabsSL2Z}) to the cross-section in dual CFT in three different pictures, that is defined by
\cite{Malda-Strom}
\be
P_{abs} \sim {T _L}^{2h_L - 1} {T _R}^{2h_R - 1} \sinh \left( {\frac{{{\tilde{\omega}} _L }}{{2{T _L} }} + \frac{{{\tilde{\omega}} _R }}{{2{T _R} }}} \right)\left| {\Gamma \left( {h_L + i\frac{{{\tilde{\omega}} _L }}{{2\pi {T _L} }}} \right)} \right|^2 \left| {\Gamma \left( {h_R + i\frac{{{\tilde{\omega}} _R }}{{2\pi {T _R} }}} \right)} \right|^2, \label{Pabs2CFTSL2Z}
\ee
where $h_{L},h_R$ are the conformal weights. The agreement between (\ref{PabsSL2Z}) and (\ref{Pabs2CFTSL2Z}) further could be found by choosing the suitable left and right frequencies ${\tilde{\omega}} _L,{\tilde{\omega}} _R$. In order to do so, we consider the first law of thermodynamics for the general charged rotating black holes
\be \label{BHthermoLaw}
\delta S_{BH} = \f{\delta M}{T_H} - \f{\Omega _H}{T_H} \delta J - \f{\Phi _H}{T_H} \delta Q,
\ee 
where $T_H$, $\Omega_H$ and $\Phi_H$ are given by (\ref{en2}), (\ref{en3}) and (\ref{en4}), respectively.
On the other hand, we have
\be \label{delSCFT}
\delta S_{CFT} = \frac{{\delta E_L }}{{T_L }} + \frac{{\delta E_R }}{{T_R }}.
\ee 
that comes from the variation of Cardy entropy formula. We may identify $\delta M$ as $\omega$, $\delta J$ as $m$, $\delta Q$ as $q$ and $\delta E _{R,L}$ as $\tilde{\omega} _{R,L}$. When we equate (\ref{BHthermoLaw}) with (\ref{delSCFT}), we find a family of left and right frequencies
\be 
\tilde \omega _{L,R} = \omega _{L,R} - q_{L,R} \mu _{L,R} ,\label{ff}
\ee
where
\begin{eqnarray}
&&\omega _{L} = \frac{(r_+^2 - r_*^2)\left[r_+^2(\kappa^2 +1) +2(a+n)^2\right]}{2\left[r_+^2(a_2-a_1\kappa^2)+(a+n)^2(a_2 -a_1)\right]}\omega,~~~q_L = q, \label{oL} \\
&&\mu _{L} = \frac{e(r_+ -r_*)[r_+^2(\kappa^2 +1) +2(a+n)^2]}{2\left[r_+^2(a_2-a_1\kappa^2)+(a+n)^2(a_2 -a_1)\right]\Xi}, \label{muL} \\
&&\omega _{R} = \frac{r_+^2(\kappa^2 -1)\left[r_+^2 + r_*^2 +2(a+n)^2\right]\omega-2ar_+^2(\kappa^2 -1) m}{2\left[r_+^2(a_2-a_1\kappa^2)+(a+n)^2(a_2 -a_1)\right]}\omega , \label{oR} \\
&&\mu _{R} = \frac{er_+^2(r_+ + r_*)(\kappa^2 -1)}{2\left[r_+^2(a_2-a_1\kappa^2)+(a+n)^2(a_2 -a_1)\right]\Xi},~~~q_R = q. \label{muR} \
\end{eqnarray}
For the scalar fields, we have $ h_{L,R} = h $. To find the quantities in $ J $  and $Q$ pictures, we just have to set $ \alpha=1, \beta=0 $ and $ \alpha=0, \beta=1 $, respectively. From this fact, we conclude that the identifications support the existence of dual deformed CFT  to the  Kerr-Newman-NUT-AdS black holes. A similar calculation in general $Q '$ picture shows that the scattering results are given by the equations (\ref{muL})-(\ref{muR}), replacing the modular group parameters $\alpha$ and $\beta$ to $\eta$ and $\tau$, respectively, and changing $m'$ to $q'$. 

\section{The dual deformed hidden conformal symmetry for the Reissner-Nordstr{\"o}m-NUT-AdS black holes}\label{deformedRNNUTAdS}

As we mentioned before, a very interesting feature of the dual deformed hidden conformal generators for the Kerr-Newman-NUT-AdS, is that, they provide a realization of the dual conformal generators for the non-rotating Reissner-Nordstr{\"o}m-NUT-AdS black holes. 
In fact, such a realization of the dual deformed hidden conformal generators for the non-rotating black holes, was discovered for the Schwarzschild \cite{Lowe}, \cite{Bertini}, the Reissner-Nordstr{\"o}m \cite{Ghezelbashdeform}, and the Gibbons-Maeda-Garfinkle-Horowitz-Strominger black holes \cite{Ghezelbash}.

The dual deformed hidden conformal generators in $Q'$ picture, in the limit of no-rotation ($a=0$), reduce to
\begin{eqnarray}
L_{\pm}^a &=&
\f{\mathrm{e}^{\pm\frac{K(r_+ -r_*)}{2\left(\kappa r_+^2 - n^2\right)\Xi} t \mp 2\pi T_{R}\phi}}{\sqrt{\Delta}}\left[\mp\Delta\partial_{r}+\frac{er_+\left[(\kappa r_+ +r_*)-(\kappa +1)r\right]\Xi}{r_+-r_*}\partial_{\chi}\right. \nonumber\\
& +&\left. \frac{\left[r_+^2(\kappa ^2 r_+ +r_*)+(r_+ +r_*)n^2 -(r_+^2(\kappa^2 +1)+2n^2)r \right]\Xi}{r_+-r_*}\partial_{t}\right] , \label{FF}\\
L_{0}^a &=& \frac{\left[r_+^2\left(\kappa^2 +1 \right)+2n^2\right]\Xi}{\sqrt{K} (r_+ - r_ *) }\partial_{t}+\frac{er_+(\kappa +1)\Xi}{\sqrt{K}{(r_ + - r_ *) }} \partial_{\chi}, \label{LaQRN}\
\end{eqnarray}
in branch a, and to 
\begin{eqnarray}
{L}_{\pm}^b & = & \f{\mathrm{e}^{\pm\f{K(r_+ -r_*)(\kappa+1)}{2(1-\kappa)(\kappa r_+^2 - n^2)\Xi}t\mp2\pi T_{L}\phi}}{\sqrt{\Delta}}\left[\f{\left[r_+^2 (\kappa^2 r_+ -r_*)+(r_+ -r_*)n^2 -r_+^2(\kappa^2-1)r\right]\Xi}{r_+ -r_*}\partial_t \right.  \nonumber\\
&+ & \f{er_+\left[(\kappa r_+-r_*)-(\kappa-1)r\right]\Xi}{r_+ -r_*} \partial_\chi \mp\Delta\partial_{r}\Bigg{]},\label{GG}\\
{L}_{0}^b& = & \frac{r_+^2\left(\kappa^2 -1 \right)\Xi}{\sqrt{K} (r_+ - r_ *) }\partial_{t} + \f{er_+(\kappa-1)\Xi}{\sqrt{K}(r_+ -r_*)}\partial_\chi, \label{LbQRN} \
\end{eqnarray}
in branch b, respectively.  Moreover, the %
central charges and the CFT temperatures are given by 
\be
c_L = c_R = \frac{6e r_+(1 -\kappa)\left(\kappa r_+^2-n^2\right)}{K(r_+ -r_*)},\label{cGRN}
\ee
and
\be
T_R = \frac{Kr_+(r_+ -r_*)(\kappa +1)}{4\pi e\left(\kappa r_+^2 -n^2\right)},~~~ T_L = T_R\frac{r_+^2(\kappa^2+1)+2(a+n)^2}{r_+^2(1-\kappa^2)}\label{TLRGRN},
\ee
respectively, where
\begin{equation} 
K =1\textcolor{black}{+}\f{6(r_+^2+n^2)}{l^2},~~~  r_* = r_+ - \f{1}{K r_+}\left[r_+^2 -e^2 \textcolor{black}{-} \f{r_+^2(r_+^2 +6n^2)}{l^2} \textcolor{black}{+}\f{(3n^2-l^2)n^2}{l^2} \right] . \
\end{equation}
The central charges (\ref{cGRN}) and temperatures (\ref{TLRGRN}), lead to the Cardy entropy which is exactly equal to the Bekenstein-Hawking entropy for the Reissner-Nordstr{\"o}m-NUT-AdS black holes. We also note that in the limits of  $ \Lambda=0 $ and $ n=0 $, the generators (\ref{FF})-(\ref{LbQRN}) provide the dual CFT description for the non-rotating  Reissner-Nordstr{\"o}m-NUT and Reissner-Nordstr{\"o}m-AdS black holes, respectively.

\section{Conclusions}\label{conc}

\textcolor{black}{We construct different classes of dual deformed hidden conformal generator for the wave equation of a scalar field in the background of the Kerr-Newman-NUT-AdS black holes.}
We explicitly deformed the inner horizon of the black hole by a deformation parameter,  to find the deformed radial equation for the scalar fields. The deformation of the inner horizon is, in fact, a result of the back-reaction of the scalar field on the black hole geometry. Under the back-reaction, the inner horizon changes to a space-like singularity. 

We construct the dual deformed conformal generators in three different pictures for the scalar wave equation in the background of the Kerr-Newman-NUT-AdS black holes. We also show that the dual deformed conformal generators in the $J$ and $Q$ pictures are special cases of the generators in general picture with special elements of modular group. We also notice that the deformed hidden conformal generators reduces to the hidden conformal generators for the scalar fields, in the background of the Kerr-Newman-NUT-AdS black holes for a very special value of the deformation parameter.

A very unique feature of the deformed hidden conformal generators, is that, they provide a realization of the deformed hidden conformal symmetry for the scalar wave equation in the background of the non-rotating black holes.  More specifically, we construct the dual conformal generators for the scalar wave equation in the background of non-rotating Reissner-Nordstr{\"o}m-NUT-AdS black holes.

We also find the absorption cross-section of the charged scalar fields in the background of Kerr-Newman-NUT-AdS black holes. The results are in agreement with the cross-sections in two-dimensional conformal field theory in all three conformal pictures.

\textcolor{black}{We also emphasis that in this article, by any reference to the holographic duality between the rotating black hole and the CFT, we exactly mean that the wave equation of a scalar probe field, on the near horizon geometry of the rotating black hole, possesses the conformal symmetry \cite{KerrCFT1}-{\cite{Castro}}.}
\vspace*{8mm}

\noindent {\large{\bf Acknowledgments}}
\vspace*{4mm}

A. M. Ghezelbash would like to acknowledge the support, by the Natural Sciences and Engineering Research Council of Canada. 
M. F. A. R. S., A. S, and F. P. Z. are supported partly, by ``Riset PMDSU 2018" and ``PKPI Scholarship"  from Ministry of Research, Technology, and Higher Education of
the Republic of Indonesia.  \newline

\end{document}